\documentclass[a4paper,12pt,epsf]{article}
\usepackage[T1]{fontenc}
\usepackage{babel}
\usepackage{graphicx}
\usepackage{epsf}
\usepackage{cite}

\begin{document}
\begin{center}
~

\vspace{-3cm}

{\Large \bf
The KARMEN Time Anomaly: 
Search for a Neutral Particle of Mass 33.9\,MeV in Pion Decay
}  

\vspace{1mm}

\normalsize
M.~Daum$^{1,\ast}$,
M.~Janousch$^{2,\dagger}$,
P.-R.~Kettle$^{1}$,
J.~Koglin$^{1,3}$,
D.~Po\v{c}ani\'{c}$^{3}$,
J.~Schottm\"uller$^{1}$,
C.~Wigger$^{1}$,
Z.G.~Zhao$^{4}$\\
\end{center}
\noindent
$^{1}$~PSI, Paul-Scherrer-Institut, CH-5232 Villigen-PSI, Switzerland.\\
$^{2}$~IPP, Institut f\"ur Teilchenphysik, Eidgen\"ossische Technische
Hochschule 
{\hspace*{3mm}}Z\"urich, CH-5232 Villigen-PSI, Switzerland.\\
$^{3}$~Physics Department, University of Virginia, Charlottesville,
Virginia 22901, 
{\hspace*{3mm}}USA.\\
$^{4}$~IHEP, Institute of High Energy Physics, Chinese Academy of Science,\\
{\hspace*{3mm}}Beijing 100039, The People's Republic of China.\\
$^{\ast}$E-mail address: Manfred.Daum@psi.ch  \\
Tel.: +41 56 310 36 68; Fax: +41 56 310 32 94.\\
$^{\dagger}$Present address: 
Paul-Scherrer-Institut, CH-5232 Villigen-PSI, Switzerland. 
\begin{center}
{\bf ABSTRACT}
\end{center}

\vspace{-3mm}

\noindent

We have searched for the pion decay $\pi^+ \rightarrow \mu^+ X$, 
where $X$ is a neutral particle of mass 33.905\,MeV. This process was 
suggested by the KARMEN Collaboration to explain an anomaly in their 
observed time distribution of neutrino induced reactions. 
Having measured the muon 
momentum spectrum of charged pions decaying in flight, we find no evidence 
for this process and place an upper limit on the branching fraction 
$\eta \leq 6.0 \cdot 10^{-10}$ of such a decay at a 95\,\% confidence 
level.\\
{\bf PACS: 13.20.Cz, 14.60.Pq, 14.80.-j}
\newpage
\noindent
In 1995, the KARMEN Collaboration reported\cite{karmen1} on a
`long-standing' 
discrepancy between their measured and expected
time distributions of neutrino induced reactions originating from 
neutrinos produced from $\pi^{+}$- and $\mu^{+}$-decays at rest
in the primary target of the ISIS facility at the Rutherford Appleton
Laboratory. 
Further evidence of events exceeding 
the expected exponential distribution 
characterized by the muon lifetime and clustered around 3.6\,$\mu$s
after beam-on-target has since been reported\cite{karmen2}.
The speculative explanation given in Ref.\cite{karmen1} was 
that the anomalous events could originate from the rare 
pion decay process,
\begin{equation}
\pi^+ \rightarrow \mu^+ X,
\end{equation}
where {\em X} is a new massive neutral
particle.
From the time-of-flight (TOF) information and the mean flight-path 
of 17.5\,m (including 7\,m of steel shielding) the mass of $X$ 
was deduced to be 33.905\,MeV.

Various hypotheses have been put forward as to the nature of the $X$-particle.
Barger et al. \cite{barger} suggest a mainly isosinglet (sterile) neutrino
with the dominant visible decay mode
$X \rightarrow e^{+}e^{-}\nu$. 
In another paper, Choudhury and 
Sarkar \cite{cs} consider a supersymmetric solution where the 
$X$-particle is the 
lightest neutralino and decays radiatively.
Gninenko and Krasnikov\cite{gninenko} 
proposed an alternative process to explain the time anomaly by 
introducing the exotic muon decay $\mu^+ \rightarrow e^+ X$.
A recent experiment found no evidence for such a decay mode
down to a branching 
fraction of $5.7 \cdot 10^{-4}$~(90\,\% c.l.)\cite{bilger2}.

The mass of $X$ originating from decay (1) is very similar to the 
mass difference between the charged pion and the muon,
$m_{\pi^{+}}-m_{\mu^{+}}=(33.91157\pm0.00067)$MeV\cite{pdg,assamagan}.
Thus, the resulting
small Q-value makes it prohibitive to look for such a particle in 
`heavy neutrino searches' from pions decaying at rest\cite{hn1,hn2,hn3,hn4}. 
However, it has certain advantages when looking
at decays in flight.
The velocity of the decay muon is very 
similar to that of the
original pion, and hence, the momentum is well defined, 
$p_{\mu}\simeq p_{\pi}\cdot m_{\mu}/m_{\pi}$.
It also follows that the energy-loss of the muon is almost equal to 
that of the pion.
Furthermore, the flight
direction of the muon differs only slightly from that of the pion.
Thus, a pion beam-line itself can be used as a spectrometer to 
optimally separate muons of
the decay\,(1) from both other beam particles and muons from the 
normal pion decay,
\begin{equation}
\pi^+ \rightarrow \mu^+ \nu_{\mu}.
\end{equation}
In addition, the pions can also be used to adjust 
the electronics timing and counter thresholds 
for muons from decay\,(1).

Soon after the initial KARMEN publication \cite{karmen1}, 
several searches were undertaken to look for the rare 
decay\,(1)\cite{daum,bilger,bryman}.
Using the high intensity pion channel $\pi$E1 at
PSI and the decay-in-flight method, we were able to set 
the most sensitive upper limit on the 
branching fraction 
\begin{equation}
\eta =
\frac{\Gamma(\pi^+ \rightarrow \mu^+  X)}{\Gamma(\pi^+ \rightarrow \mu^+ 
\nu_{\mu})}
\end{equation}
of this decay at
$\eta \leq 2.6 \cdot 10^{-8} $ (95\%~c.l.) \cite{daum}. 
In order to improve on this sensitivity, we performed a new experiment
using a similar method with the following
improvements:
(i) a better phase space definition of the pion beam;
(ii) a better phase space definition in the muon spectrometer;
(iii) additional background suppression by using
a new beam-line optics and an improved vacuum system
with active and passive collimators.

Our new experimental setup, shown in Fig.\,1, consists of three
main parts.
The first part is the pion transport system from the pion production 
target to the end of the dipole magnet AEF51.
Here, the collimators FSH51, FSH52, FS53, and ZBN were used to define 
the pion phase space centred at 150\,MeV/c 
with a momentum spread of 1.2\,\% (95\,\%).
This central beam momentum was chosen so that
the TOF-information could be used to optimally separate pions
from positrons and muons also contained in the beam. 
The second part, between the dipole magnets AEF51 and ASL52, 
serves as the 4.5\,m long field-free pion-decay region in which about 
42\,\% of the entering pions decay. With the proton accelerator
operating at a beam current of 1.5\,mA, the pion decay rate 
was approximately 10$^7$\,Hz in this region.
The active collimators V1--V4 were positioned within the vacuum system
to suppress the background of scattered particles originating from
both before and within the decay region. The final section,
from the dipole magnet ASL52 to the counter S3, is the momentum analyzing 
spectrometer. The active collimators V0 and V5 were used together with 
the beam counters S1--S3 and the passive collimator FS54 to define 
the spectrometer acceptance such that only particles with a momentum spread 
of 2.3\,\% and divergences similar to those of the original pions were 
accepted. For normalization purposes, a muon telescope in the decay 
region was used together with a proton monitor in the primary
proton beam.

The experimental method involved firstly tuning the entire beam-line
according to beam transport calculations
using 150\,MeV/c pions. These pions were then used to set up the electronics 
timing and counter thresholds. 
The triple coincidence S1$\cdot$S2$\cdot$S3 of the beam counters 
defined our trigger events.
In a next step, `decay muon scans' were performed by leaving
the magnets of the pion transport
system at their original 150\,MeV/c settings while the 
spectrometer magnets were scaled in steps of 0.5\,MeV/c from
103.0\,MeV/c to 124.0\,MeV/c.
In this momentum range, muons from the normal decay (2) 
of 150\,MeV/c pions have 
decay angles of about 255\,mrad
and are not accepted by the beam-line and the spectrometer.
Muons from decay (1), however, have a maximal decay angle of about 5\,mrad
and are accepted by the spectrometer. The scans
were performed over this broad momentum range in order to characterize the 
background distribution over which excess events centred at 
$p_{\mu} \simeq p_{\pi} \cdot m_{\mu}/m_{\pi}$
(i.e. 113.5\,MeV/c)
would indicate evidence for the existence of decay (1).

The shape and position of the peak which would be produced
from decay (1) were deduced 
from `pion scans' and `muon scans'. The `pion scans' were performed
by scanning the spectrometer magnets around 150\,MeV/c, while the 
`muon scans' were obtained by firstly scaling the magnet currents in 
the pion transport system by a factor of $m_{\mu}/m_{\pi}=0.757$
(i.e. to a momentum of 113.5\,MeV/c) and then
scanning the spectrometer magnets around this momentum.
These scans were also used to test the magnet scaling properties.
Deviations from linearity were found to be 
negligible for all but one magnet.
Here, the deviation was 1.2\,\% and was accounted for in acceptance 
calculations and magnet setting corrections.

Our data-taking procedure consisted of `decay muon scans' in which 
an automated procedure was followed ensuring that for each of the scans
the momentum settings were reproducible\cite{koglin}.
At each momentum setting, the pulse-height and timing information
of all beam and active veto-counters were recorded
event-by-event for a given number of protons on the 
pion production target. 
To demonstrate our particle identification ability,
two-dimensional TOF-distributions
for events which have passed initial timing and 
veto cuts are shown in Fig.\,2. 
One can clearly distinguish pions, contained in the 
indicated timing box, from muons and positrons also present in the 
beam.
The candidate muon events from 
decay\,(1) 
are expected to be located in plot\,(b)
at the same location as the pions in plot\,(a). 
These muons are clearly distinguishable from scattered muons as 
well as scattered positrons and positrons from other decay sources.

Based on modifications that were made to our 
spectrometer acceptance during the beam time,
the data can be divided into four sets.
For each scan, the momentum spectrum of muons was obtained 
from the number of candidate events at each momentum setting of the
spectrometer. These candidate 
events had to fulfil the following timing and veto
conditions: (i) their timing must be consistent with that of 
muons having been produced from pions in the decay region and travelling
with the velocity corresponding to the spectrometer settings;
(ii) they must not be accompanied by a corresponding hit in any of the 
22 veto counters. 
With the veto cuts, more than 90\,\% of the events passing the timing cuts
were rejected.
The data were then normalized to the number of pion decays in the 
decay region.
The momentum spectrum of one of the scans is shown in Fig.\,3.

In order to model the background, the central points of the
momentum distribution were excluded while the remaining points were fitted with 
various functions such as second and fourth order polynomials.
The best fits were obtained with a general hyperbola which has five 
degrees of freedom. 
Each scan was then fitted 
with a Gaussian function simulating
the expected distribution of muons from the hypothetical decay (1)
together with the background function.
The width and the position of the Gaussian function were
determined from computer simulations\cite{koglin,turtle} 
using the pion and muon scan data discussed above.
The height of the Gaussian together with the five parameters
of the hyperbolic background function were left as free fit parameters.
The branching fraction $\eta$, see eq.\,(3), was determined by weighting
the fitted height of the Gaussian with the spectrometer
acceptance which was determined from pion scans and 
beam transport simulations\cite{koglin,turtle}.

The fit to one of our scans is shown in Fig.\,3 along with the 
effect one would expect to see for a branching fraction of 
$\eta = 5 \cdot 10^{-9}$. 
No indication
for the existence of the hypothetical decay (1) is evident in 
our data, cf.\ also Fig.\,4.
The expected level of background events originating
from the radiative pion decay, $\pi^+ \rightarrow \mu^+ \nu_{\mu} \gamma$,
was obtained from a Monte Carlo simulation and is also indicated in Fig.\,3.
The remaining 
background is assumed to originate predominantly
from scattered particles and small inefficiencies in the 22 veto
counters.

The fit results to $\eta$ for our 28 different scans are displayed in
Fig.\,4. The combined confidence level of the fits to the 
28 scans is 29\,\%\cite{koglin}.
The weighted mean is $\eta = (1.27 \pm 2.27)\cdot 10^{-10}$.
Our systematic uncertainty was found to be 5\,\% and accounts for 
spectrometer acceptance and the overall normalization
uncertainties\cite{koglin}.
Therefore, in order to obtain a conservative result, 
a factor of 1.05 was applied multiplicatively to both $\eta$ and 
its uncertainty which yields the branching fraction
\begin{equation}
\eta = (1.3 \pm 2.4) \cdot 10^{-10}.
\end{equation}
Following the Frequentist's approach described 
in Ref.\,\cite{feldman,pdg}, we find an upper limit of
\begin{equation}
\eta \leq 6.0 \cdot 10^{-10} \mbox{ (c.l. = 95 \%).}
\end{equation}
This new upper limit is a factor of 45 lower than
the previous limit\cite{daum,pdg}; note that our sensitivity [cf.\ eq.\,(4)], 
however, was increased by two orders of magnitude.
The new result rules out the supersymmetric 
solution of the KARMEN anomaly considered by Choudhury and 
Sarkar\cite{cs}. Other solutions in which the
$X$-particle is considered as an isosinglet (i.e.~sterile) neutrino
consistent with the KARMEN hypothesis are still possible
down to a level of about 10$^{-13}$\cite{barger,govaerts}.
In a recent experiment\,\cite{nutev}, upper limits for $\eta$ were obtained
between 10$^{-13}$ and $10^{-6}$ for lifetimes 
$10^{-10}$s$\leq \tau_X \leq 10^{-1}$s. 

This project was supported by the 
Schweizerischer Nationalfonds and the US 
National Science Foundations.

\newpage
\noindent
\subsection*{Figure Captions}
\noindent
Figure\,1:\\
The beam-line consists of three parts:
(i) the pion transport system up to and including the magnet AEF51;
(ii) the pion decay region between the magnets AEF51 and ASL52;
(iii) the muon spectrometer from the magnet ASL52
to the beam counter S3. 
ASZ, ASY, ASL, AEF: dipole magnets; QTH, QTB, QSL, QSE, QSK: 
quadrupole magnets; FSH, FS, ZBN: passive collimators;
S1--S3: beam scintillation counters; V0--V5: active (scintillation counters)
and passive (lead jaws) collimators.
~\\
~\\
Figure\,2:\\
Experimental TOF-distributions with the relevant background event
types ($^*$~refers to scattered particles). The upper plot\,(a) shows 
the situation when the whole beam-line is
set to accept pions of 150\,MeV/c. Plot\,(b) shows the 
situation for the central momentum of a `decay muon scan' 
(113.5\,MeV/c). The horizontal axes indicate 
the TOF of particles from the
production target to the counter S1 relative to the 
accelerator time structure of 19.75\,ns.
The vertical axes indicate the TOF between the 
counters S1 and S3. 
The events in the box in (b) have the same timing as pions in (a),
and thus may contain signal as well as background events.
It should be noted that the 
number of protons for which the data were recorded is 
seven orders of magnitude higher in (b) than in (a).
~\\
~\\
\noindent
Figure\,3:\\
Experimental data of one `decay muon scan'.
The solid curve is the fit to the data (hyperbola plus Gaussian);
the $\chi^2$ is 33.6 for 37 degrees of freedom. 
The shaded peak indicates the expected effect of the hypothetical decay (1) 
for $\eta = 5.0\cdot$10$^{-9}$. The shaded band 
shows the expected level of background from the 
radiative pion decay, $\pi^+ \rightarrow \mu^+ \nu_{\mu} \gamma$,
within computational uncertainties ($\pm 1 \sigma$).
The minimum in the event distribution around 113.5 MeV/c arises from the fact
that at this momentum, muons from the normal pion decay (2) have 
a maximal emission angle and thus a minimal chance to enter the acceptance 
of the spectrometer. 
~\\
~\\
\noindent
Figure\,4:\\
Results of the fits to the branching fraction $\eta$ for our 28 scans.
The weighted mean for $\eta$ is indicated by the solid line with the
total uncertainty ($\pm 1 \sigma$) given by the dashed lines.

\begin{figure}[htbp]

\vspace{-1cm}

{\bf \caption{~}}

\vspace{1cm}

\begin{center}

\epsfxsize=8.6cm
\epsfbox{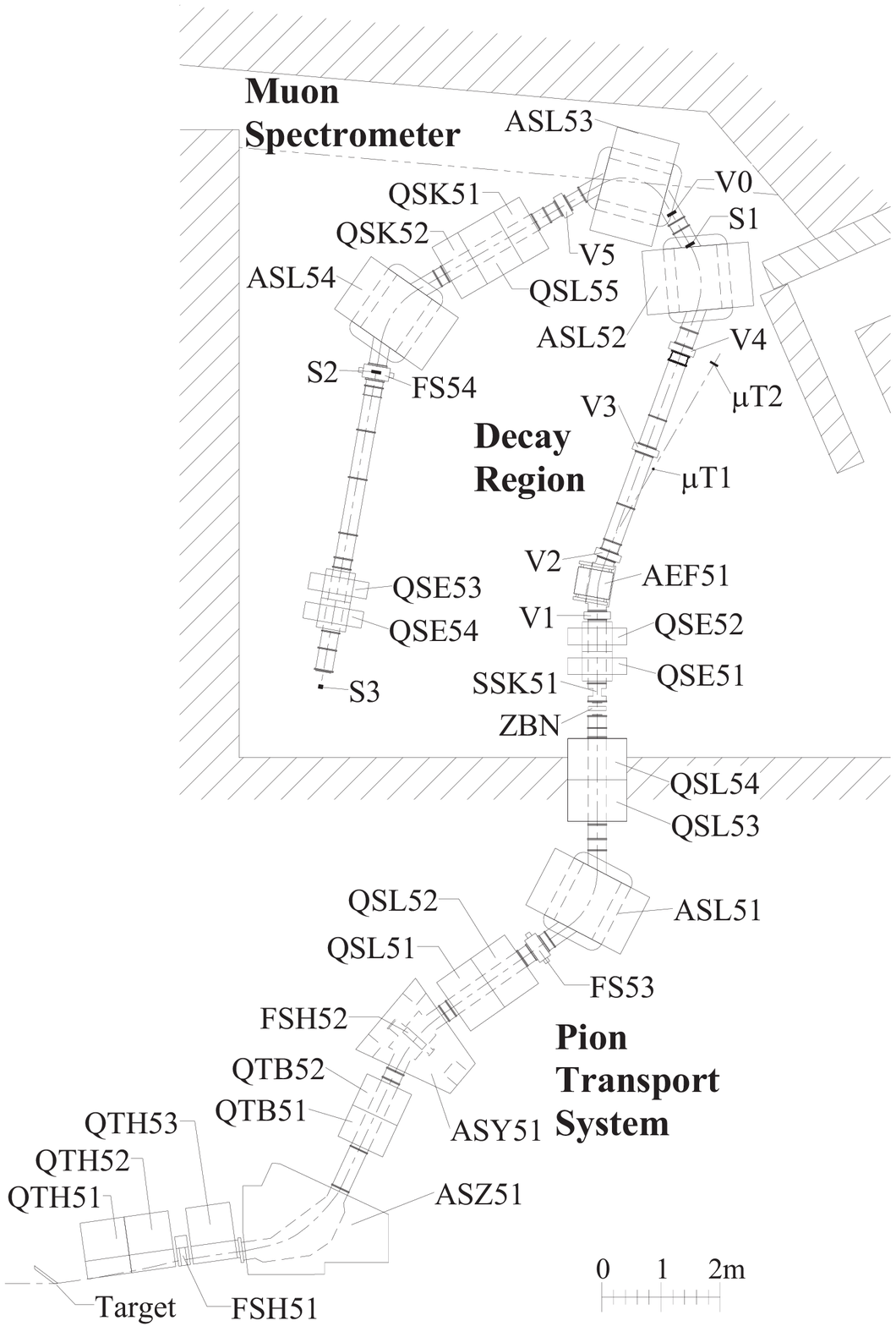}
\label{fi:setup}
\end{center}
\end{figure}

\begin{center}
\begin{figure}[htbp]

\vspace{-1cm}

{\bf \caption{~}}

\vspace{1cm}

\epsfxsize=8.6cm
\epsfbox{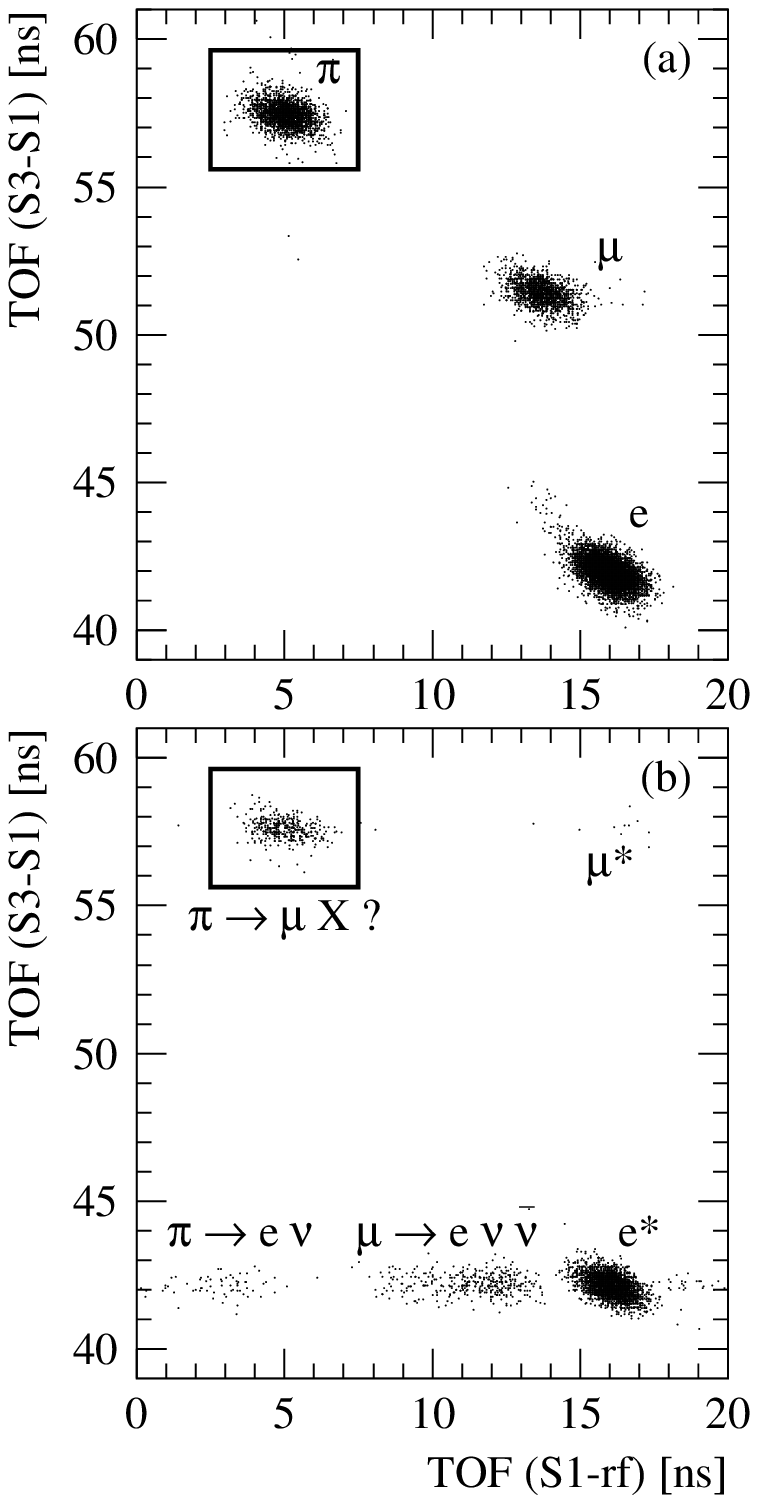}
\end{figure}
\end{center}

\begin{center}
\begin{figure}[htbp]

\vspace{-1cm}

{\bf \caption{~}}

\vspace{1cm}

\epsfxsize=8.6cm
\epsfbox{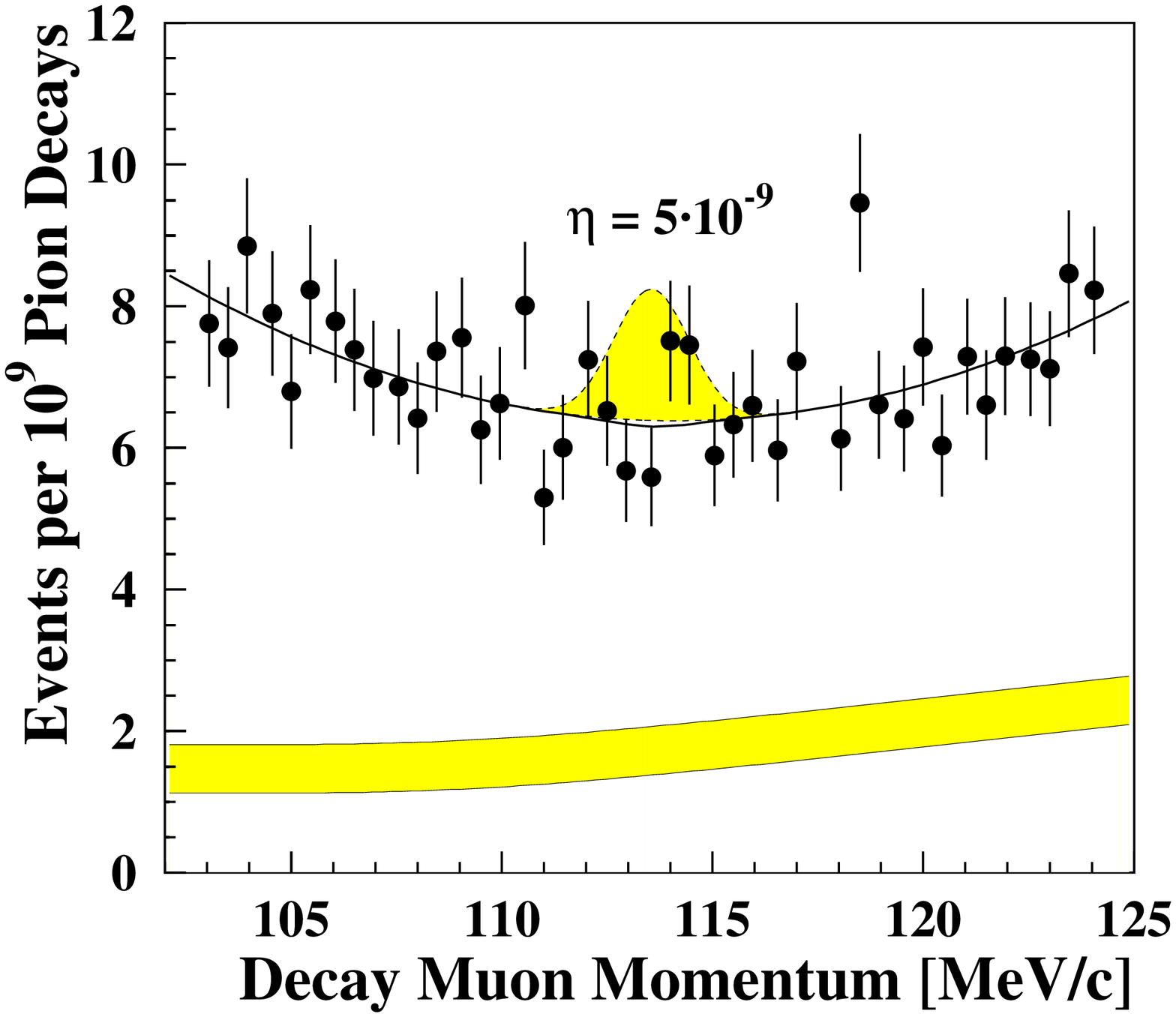}
\end{figure}
\end{center}

\vspace{10cm}

\begin{center}
\begin{figure}[htbp]

\vspace{-1cm}

{\bf \caption{~}}

\vspace{1cm}

\epsfxsize=8.6cm
\epsfbox{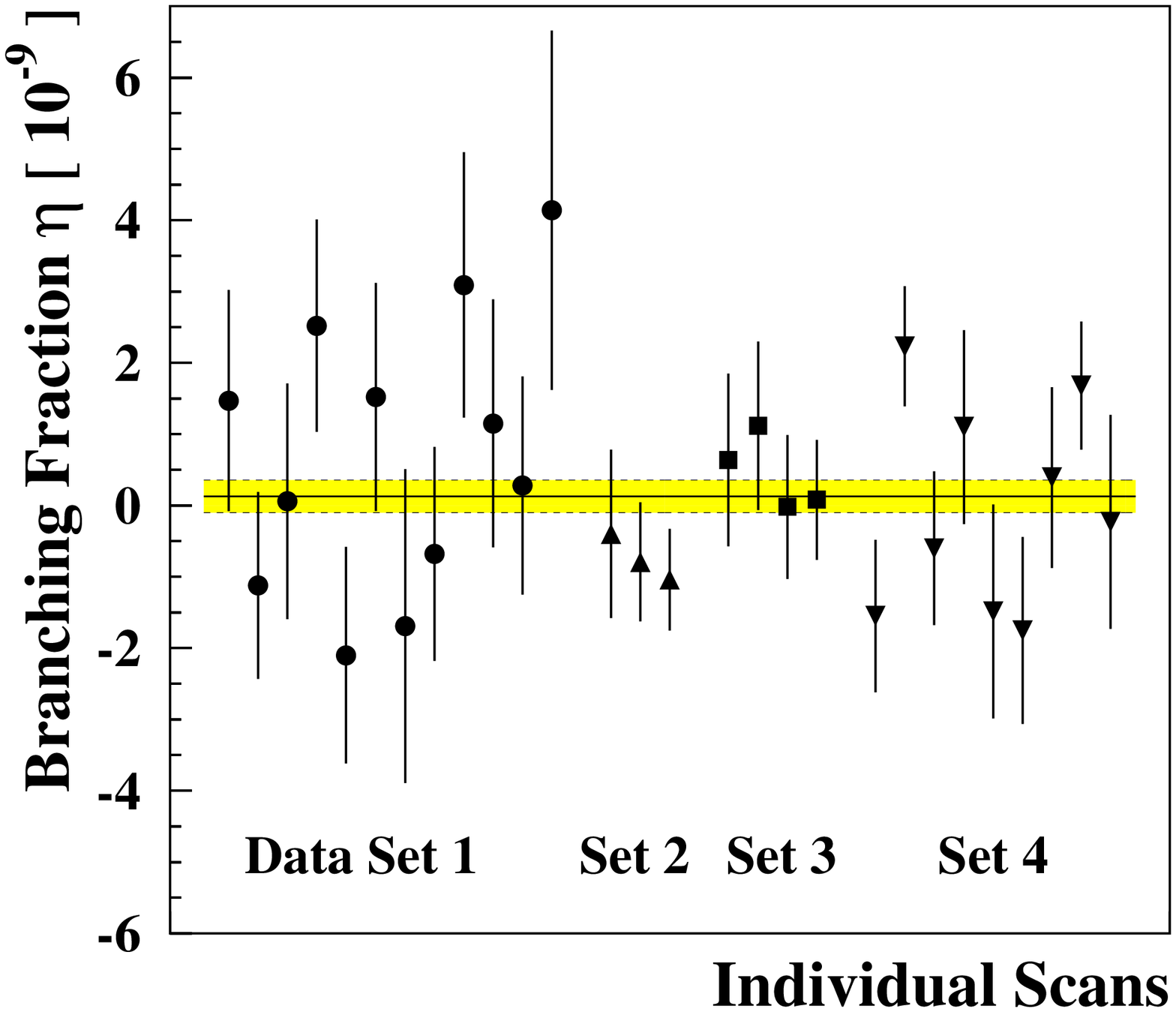}
\end{figure}
\end{center}

\end{document}